\documentstyle [12pt]{article}
\newcommand{\beq}{\begin{equation}}

\newcommand{\eeq}{\end{equation}}

\newcommand{\la}{\lambda}
\newcommand{\si}{\sigma}

\newcommand{\beqa}{\begin{eqnarray}}
\newcommand{\eeqa}{\end{eqnarray}}
\begin{document}
\hbox{\hskip 4.0 true in ITP-SB-92-70}
\hbox{\hskip 4.0 true in CEBAF-TH-92-32}
\vbox{\vskip 0.6 true in}

\centerline{QCD SUM RULES AND COMPTON SCATTERING}
\vbox {\vskip 0.5 true in}

\centerline{Claudio Corian\`{o}$^{1,4*}$, Anatoly Radyushkin$^{2,3**}$
 and George Sterman${^1}$}
\vskip 0.4in
\centerline { ${^1}$Institute for Theoretical Physics, State University of
New York,}
\centerline{Stony Brook, New York 11794-3840}
\centerline{ ${^2}$ Physics Department, Old Dominion University,
Norfolk, VA  23529}
\centerline{ ${^3}$ Continuous Electron Beam Accelerator Facility,
Newport News, VA 23606}
\centerline{ ${^4}$ Department of Physics, University of Stockholm}
\centerline{S 113 Vanadisvagen, Stockholm, Sweden}
\vbox {\vskip 0.5 true in}

\centerline {ABSTRACT}
\medskip
\narrower{We extend QCD sum rule analysis to
moderate energy fixed angle Compton scattering.
In this kinematic region there
is a strong similarity to the sum rule treatment of electromagnetic
form factors, although the four-point amplitude requires a
modification of the Borel transform.
To illustrate our method, we derive the sum rules for
 helicity amplitudes in pion
 Compton scattering and estimate their large-$t$ behavior in the local
duality approximation.

\vbox{\vskip 1.0 true in}

\noindent $^{(*)}$ {\footnotesize A. Della Riccia fellow} \\
$^{(**)}$ {\footnotesize On leave from the Laboratory of Theoretical
Physics, JINR, Dubna, Russian Federation}

\newpage
\section{Introduction}

The method of QCD sum rules \cite {SVZ}
originated from an analysis of the high energy behavior of two-point
correlation functions of QCD currents.
More than 10 years ago, this method was
 extended to elastic form factors \cite{IS,NR1}
by treating three-point functions.
In the sum rule method, one considers Green functions of interpolating
fields given by
current-correlators which have nonzero resonant contribution on
selected lowest states. The non-perturbative resonant region is then
enhanced over the continuum by using  Borel transforms \cite{SVZ,IS,NR1}.
A problem in the more
general cases \cite{BI} involving four or higher-point functions is the larger
number of invariants on which the
function depends.
In this paper, our goal is to extend
the QCD sum rule approach
to fixed angle pion Compton scattering, a process
described by an amplitude
which is closely related to the pion  electromagnetic form factor.

Our considerations build on previous work on sum
rules for the pion form factor in Refs.\ \cite{IS} and \cite{NR1},
and are made possible by the observation that, at high energy and fixed
angle, the underlying quark-photon scattering subprocess
has a short-distance nature.

The approach that we investigate is also related to the observation
 (see, {\it e.g.,} \cite{ES}) that QCD sum rules can be formulated as
a specific variant of the finite-energy sum rules
analogous  to those applied originally to high-energy hadron-hadron scattering
\cite{LST,S,DAFFR}.
Compared to the sum rule treatment of form factors \cite{IS,NR1},
we must take into account the presence of additional
cuts in the $u$-channel
when we construct a dispersion relation.
These additional cuts force us to work in  a kinematical regime in $s$ and $t$
where
the cuts are far from the region enclosed by the
integration contour.
Our aim is to show that if we choose
$s$, $t$ and $s+t$
to be  moderately large
in the physical region
$s>0, t<0$, then
a dispersion relation can be written down and, hence, a
QCD sum rule
can be derived.  These sum rules result in invariant
amplitudes with $s$ and $t$ dependence in general.  This
dependence is not expected to be the same as for perturbative
calculations based on elastic scattering factorization theorems \cite{pQCDff},
because the sum rule calculation begins at $O(\alpha_s^0)$.
Thus, when our method is extended to the proton, it may be
possible to confront both the sum rule and the standard
perturbative calculations with experiment at a fixed energy.
This would help clarify some controversial issues concerning
the relative importance of these contributions at
available energies \cite{Radyconf,Isgur,Li}.

Our arguments are illustrated by an explicit calculation of the
discontinuity in the perturbative amplitude at the lowest order,
and by a derivation of the sum rules for
invariant
amplitudes in the case of Compton scattering of the pion.

In Section 2, we
reformulate the QCD sum rule approach to   the pion
electromagnetic form factor.
To this end, we modify the standard procedure of Refs. \cite{SVZ,IS,NR1}  by
employing an integral
representation of the Borel transform.
In Section 3, we develop the relevant dispersion relation for the
four-point amplitudes, and derive a sum rule by using
a variation
of the Borel transform.
In Section 4, we discuss the isolation of the pion contribution.
Section 5 contains our results
for the lowest order
spectral functions and the
QCD sum rule estimate
for the
Compton amplitude in the ``local duality"
approximation.
Our conclusions are summarized in Section 6.

\section{\bf  Form Factor Sum Rules }

The most popular choice of the
interpolating fields
in the pion case is the axial current
\beq
\eta_\mu =\bar{u}(x)\gamma_\mu \gamma_5 d(x),
\eeq
whose projection onto a single-pion state $|p \rangle$
is given by
\beq
\langle 0|\eta_{\mu}(x)|p \rangle =i f_\pi p_{\mu}e^{-i {p\cdot x}}.
\eeq
The electromagnetic current $J_{\mu}$ defining the pion form factor is given by
\beq
J_\mu =\frac2{3}\bar{u}\gamma_\mu u - \frac1{3} \bar{d}\gamma_\mu d.
\eeq

To study the form factor, one  analyzes the 3-point correlation function
of the electromagnetic and two axial currents \cite{IS,NR1}

\beq
\Gamma_{\mu\nu\sigma}(p_1,p_2)
=   \int  \langle 0|T\left(\eta_{\sigma} (y)J_{\nu}(0)
\eta^\dagger_{\mu}(x)\right)|0 \rangle e^{-i p_1 x +i p_2 y} d^4 x d^4 y .
\label{gamma3}
\eeq

Analogously, to investigate pion Compton scattering, one
should consider the 4-point function involving two axial and two
electromagnetic currents

\beqa
\Gamma_{\nu\la\mu\si}(p_1,p_2,s,t)
&=& i\int d^4 x d^4 y d^4 z e^{-i p_1\cdot x +i p_2\cdot y - i q_1\cdot z}
\nonumber \\
&\ &\ \times \langle 0|T\left(\eta_{\sigma}
(y)J_{\nu}(z)J_\lambda(0)\eta^{\dag}_\mu (x)\right)|0 \rangle ,
\label{gamma4}
\eeqa
where
the photon lines are assumed to be on-shell.

Let $T_3(p_1^2,p_2^2,t)$,
with $t= (p_2-p_1)^2$, denote one of the scalar invariant
amplitudes
isolated from the tensor expansion of $\Gamma_{\mu\nu\la}$
\cite{IS,NR1}. The amplitude
$ T_3$ obeys the double dispersion relation
\beq
T_3 (p_1^2,p_2^2,t)= \int_{0}^{\infty}ds_1 \int_{0}^{\infty}ds_2\,
{\Delta_3 (s_1,s_2,t)\over (s_1-p_1^2)(s_2-p_2^2)} + \,\,
{\rm subtractions},
\label{dispers}
\eeq
where $\Delta_3 (s_1,s_2,t)$ is the double discontinuity of $T_3$
across the branch cut associated with thresholds in $p_1^2$ and
$p_2^2$.

The next step within the QCD sum rule approach is to introduce
the double ``Borel'' transform  ${\Phi_3}(M_1^2,M_2^2,t)$ \cite {IS,NR1}
of $T_3 (p_1^2,p_2^2,t)$.  The Borel transformation can be defined in different
ways. For our purposes, the most convenient is the
integral representation
\beq
{\Phi_3}(M_1^2,M_2^2,t)=
{1 \over (2\pi i)^2}
\int_{C}{dp_1^2\over M_1^2}\int _{C}
{dp_2^2\over M_2^2} \, e^{-p_1^2/M_1^2 - p_2^2/M_2^2}
T_3(p_1^2,p_2^2,t),
\label{borel}
\eeq
where $C$ denotes any contour that encloses the
  branch cuts in the $p_1^2$ and $p_2^2$ planes, from zero to infinity.
For any non-zero values of the $M_i^2$, the contours $C$ may be thought of as
closed across the cut at positive infinity, where the integrand is
exponentially suppressed.

For the 3-point function, the integral form (\ref{borel}) is completely
equivalent to the standard procedure  \cite{SVZ,IS,NR1} for
constructing the Borel transform.  However, as we will see, it is very
suggestive of an extension to the four-point function related to
fixed-angle high-energy pion Compton scattering.

Note that eq.~(\ref{borel})  immediately gives zero
result
for the subtraction terms
in $T_3 (p_1^2,p_2^2,t)$, since they are polynomials in the $p_i^2$.
Applying (\ref{borel}) to (\ref{dispers}) then gives \cite {IS,NR1}
\beq
\Phi_3 (M_1^2,M_2^2,t)
= \int_{0}^{\infty}{ds_1\over M_1^2}
\int_{0}^{\infty} {ds_2\over M_2^2}\, \Delta_3
(s_1,s_2,t)
e^{-s_1/M_1^2 -s_2/M_2^2}.
\label{phi3}
\eeq

To derive a sum rule, we employ
our freedom in choosing the original contours
$C$ in the definition of the double transform,
eq.\ (\ref{borel}).
Two relevant choices are shown in Fig. 1.
Contour $ C_a$ encloses the branch cut closely, while $ C_b$
encloses it with a circle of radius $\zeta^2$ about the origin, and joins
with $C_a$ for $|p^2|> \zeta^2$.
Let us suppose that we choose
\beq
 \zeta^2 \gg \mu^2
\label{zeta}
\eeq
where
$\mu$
is the scale associated with nonperturbative power
corrections in $T_3(p_1^2,p_2^2,t)$.
The inequality makes it reasonable to apply the operator product expansion
along $C_b$. Around the contour $C_b$, we may now approximate
\beq
T_3 (p_1^2,p_2^2,t)=T_3^{pert}(p_1^2,p_2^2, t)\ + \ {\rm power
\ corrections},
\label{Cb}
\eeq
where ${T_3}^{pert}(p_1^2,p_2^2,t)$ is the perturbative expansion for $T_3$
starting with the diagrams of Fig. 2.
The power corrections indicated in eq.~(\ref{Cb}) are proportional to the
familiar
vacuum condensates,
\beq
{\rm power\  corrections} = K_1 {\langle 0|\alpha_s
G_{\mu\nu}G^{\mu\nu}|0 \rangle \over (p_1^2)^2} + K_2{\langle
0|\alpha_s (\bar{q}q)^2|0 \rangle \over (p_1^2)^3} +...,
\label{conden}
\eeq
with $K_1$ and $K_2$
being, in general,
functions of $t/p_1^2$ and $p_1^2/p_2^2$ .
We shall assume that $\zeta^2 =O(t)$ in eq.\ (\ref{zeta}),
so that
possible higher-order
logarithmic corrections in these variables are not large.

Around the contour $C_a$, the approximation of eq.~(\ref{Cb})
is clearly inappropriate.
Instead, for small values of
$p_1^2$ and $p_2^2$, we parameterize $T_3$ phenomenologically - in
terms of dominant resonances plus continuum:
\beq
\Delta_3(p_1^2,p_2^2,t)=\Delta_3^{phen}(p_1^2,p_2^2,t).
\label{phen}
\eeq
The sum rules now result from eq.~(\ref{phi3}) by equating the two
transforms, one around
$C_a$
and one around $C_b$, using the two approximations, eqs. (\ref{Cb}) and
(\ref{phen})
respectively, for  $\Delta_3,$
\beqa
0 &=&\int_{0}^{\infty}{ds_1\over M_1^2}\int_{0}^{\infty}{ds_2\over
M_2^2}(\Delta_3^{pert}-\Delta_3^{phen}) e^{-s_1/M_1^2 -
s_2/M_2^2}\nonumber \\
&\ &\hbox{\hskip 0.4 true in} +\ {\rm power\ corrections},
\label{3sr}
\eeqa
with ${\Delta_3}^{pert}$ the double discontinuity of ${T_3}^{pert}$ around the
$p_1^2$ and the $p_2^2$ cuts.
Here the contributions of power corrections will appear only from the
(counterclockwise)
circle of radius
$\zeta^2$
for $C_a$ in Fig. 1,
\beq
{1\over 2\pi i}\int_{|p^2|=
\zeta^2}
{dp^2\over M^2}{e^{-p^2/M^2}\over
(p^2)^n}={(-1)^n\over (n-1)! (M^2)^n}.
\eeq
The remainder of the $C_a$ integral gives zero for power corrections,
since they are well approximated by functions that are analytic
except at the origin. Of course, this is an approximation, but an
adequate one for
$\zeta^2 \gg \mu^2$,
as specified in eq.~(\ref{zeta}).
The power corrections have been computed in Refs.~\cite{IS,NR1}.
Once the power corrections are specified in eq.~(\ref{3sr}),
information can be extracted
when an explicit form is given for ${\Delta_3}^{phen}$.
For the form factor, the standard
choice
``first resonance plus continuum''
is made in Refs.~\cite{IS,NR1}:
\beqa
\Delta^{phen}(s_1,s_2,t)
 &=&  T_3^{\pi}(t)\delta (s_1 - {m_{\pi}}^2)\delta (s_2- {m_\pi}^2)
 \nonumber \\
&\ &\ +\  {\Delta_3}^{pert}(s_1,s_2,t)(1-\theta (s_0-s_1)\theta(s_0-s_2)),
\label{cont}
\eeqa
where $T_3^\pi (t)$ is the pionic scalar amplitude and
$s_0$ is a parameter
specifying, in this model, the onset of  continuum.
Using (\ref{cont}) in (\ref{3sr}), we find the $M^2$- and $s_0$- dependent sum
rule,
\beqa
T_3^{\pi}(t) &=& \int_{0}^{s_0}ds_1
\int_{0}^{s_0}ds_2\ \Delta_3^{pert}(s_1,s_2,t)e^{-(s_1 +
s_2)/M^2} \nonumber \\
&\ &\hbox{\hskip 0.4 true in} +\ {\rm power \,\, corrections},
\label{t3pi}
\eeqa
where we have set $M_1^2=M_2^2=M^2$ and
neglected $m_\pi^2$ compared to
all other scales.

Note that
contributions from the states beyond $\sqrt{s_0}$ in mass have cancelled here.
Fixing $s_0$ from the requirement of the best stability of the
right hand side
of eq.~(\ref{t3pi}) with respect to variations of the Borel parameter
$M^2$ (this gives $s_0\approx$ 0.7 $GeV^2$) and then taking the limit
$M^2 \to \infty$ (which eliminates power corrections), one  arrives at
the ``local duality" formula for the pion
form factor \cite{NR1}.
We now turn to the application of this reasoning to the four-point amplitude.

\section{\bf Sum Rules for Four-Point Amplitudes}

In general, the analytic structure of a four particle scattering
amplitude, or a four-current correlation function, is much more
complicated than that of a three-point function.  Nevertheless,
at high energy and momentum transfer, the space-time structure
of the Compton scattering amplitudes simplifies somewhat, and becomes
similar to the corresponding form factor amplitudes.

Consider, for instance, the four-current Green function
$\Gamma_{\nu\la\mu\si}$, eq. (\ref{gamma4}).
The aim of the sum rule approach is to derive information
about $\Gamma_{\nu\lambda\mu\sigma}$ at $p_i^2=m_\pi^2$,
where
this function
develops bound-state poles.
Of course, we cannot compute $\Gamma$ perturbatively
at these momenta.  If, however, we can
use a dispersion
representation
to relate the Green function in this nonperturbative region to its
behavior in a region where perturbation theory and the operator product
expansion
may be applied, the road is open to apply the methods
described in the previous section.   We shall argue that it
is possible to do this quite directly when the Mandelstam
invariants $s=(p_1+q_1)^2$ and $t=(p_1-p_2)^2$ in
eq.~(\ref{gamma4}) are large, with the ratio $s/t$ fixed.

The first step in this process is to note that
even for on-shell external particles, the underlying hard
scattering at fixed $s/t$ remains a short-distance process
in any two-to-two scattering amplitude \cite {Akhoury}.
Perturbation theory runs into trouble from virtual particles that
either (1) are on-shell and parallel to one of the four
external momenta (collinear lines), or (2) have vanishing momenta in
all four components (soft lines).  Momentum configurations of
this type will typically diverge in the zero-mass limit,
or, equivalently, suffer from large logarithms of ratios of external
energies to masses. The situation for Compton scattering at lowest
order in electromagnetism, however, simplifies when the external photons
carry physical polarizations.  In this case, virtual lines
parallel to the incoming or outgoing photons do $not$ produce
large logarithms, nor do soft lines attached to such
collinear lines.  The failure of perturbation theory, then,
results from the lines that are collinear to one of the
hadron momenta and/or from the soft quanta that connect them.
This is qualitatively the same situation as for the elastic
form factor.  Indeed, in space-time, the incoming photon and
the outgoing photon originate at the same point.  This
observation has been used elsewhere in the context of a
perturbative QCD calculation of Compton scattering to show that
the phases of this amplitude are free of infrared divergences
\cite {FSZ}.

To take advantage of these observations, we shall need to develop
the relevant dispersion relation.  Its form will be slightly
different from the
form factor
case,
requiring the modified version of the Borel transform mentioned above.
Let us describe how this works.

Let $T_4 (p_1^2,p_2^2,s,t)$ denote a particular scalar amplitude derived from
$\Gamma_{\nu\la\mu\si}$ in eq.~(\ref{gamma4}).
We will give a specific example below.
Momentum assignments are illustrated in Fig. 3.
We will be interested in $s$ and $t$ large compared to $m_\pi^2$,
with $-t/s$ fixed but not equal to 1 or 0. We take $q_1^2=q_2^2=0$,
appropriate to physical photons.

In general, we are not able to write for $T_4$ a double
dispersion integral of precisely the form of eq.~(\ref{dispers}).
This is because, when $p_1^2$ or $p_2^2$ becomes large enough, the
kinematic identity
\beq
s+t+u= p_1^2 + p_2^2
\eeq
forces $u$ to threshold for fixed $s$ and $t$. This means, first,
that branch points in $u$ appear at $p_2^2-$dependent positions in the
$p_1^2$ plane (and vice-versa).
Second, it means that, even at large $p_1^2 +p_2^2$, a
phenomenological approximation like (\ref{cont}) would need to include
resonances in the $u$-channel.
Fortunately, this is not the end of the story. It is still possible to
write a useful double dispersion relation for $T_4$. This relation
employs the contour $\gamma$  shown in Fig. 4.

Instead of running to infinity as in eq.~(\ref{dispers}), $\gamma$  is closed
in
a circle  of radius $\lambda^2$, where $\lambda^2$ is a constant which
we may take to satisfy
\beq
\lambda^2 = (s+t)/4.
\label{lambda}
\eeq
With this choice, the $u$-channel threshold never comes closer than $(s +
t)/2$ to $\gamma$ in the $p_1^2$ plane, when $p_2^2$ is chosen to run
along the corresponding contour in its complex plane.
We therefore have, in place of eq.~(\ref{dispers}),
\beq
T_4(p_1^2,p_2^2,s,t)=
{1\over (2\pi i)^2}\int_{\gamma}ds_1\int_{\gamma}ds_2
{T_4(s_1,s_2,s,t)\over (s_1-p_1^2)(s_2-p_2^2)}.
\label{t4}
\eeq
We note that the portion of $\gamma$ that runs along the real axis
gives
an integral over a discontinuity, as in (\ref{dispers}), but,
generally speaking,
the discontinuity is {\em not}
purely imaginary.
Hence, the amplitude we are interested in will be complex in general,
even for moderate negative $p_1^2$ and $p_2^2$.
It remains, however, an analytic function of $s$ if we are far from
its thresholds.

More formal arguments for the validity of this dispersion relation,
and for the applicability of perturbation theory to the amplitude
at large values of $|p_i^2|$ are given in Appendix A.  Here we shall
assume that these results hold, and show how they lead to the
desired sum rule.

To derive a sum rule, we would like to apply the same Borel transform,
eq.~(\ref{borel}) to $T_4$ in eq.~(\ref{t4}) that we applied to $T_3$
in eq.~(\ref{dispers}).  The contours $C$ in eq.~(\ref{borel}),
however, enclose the entire real axis, and would hence cross the
contour $\gamma$ of eq.~(\ref{t4}).  This makes eq.~(\ref{borel})
unsuitable as it stands.
This technical difficulty can be circumvented if we choose a slightly
modified version of the transform.
As shown in Fig.~4, let $\Gamma$ be
any counterclockwise contour, enclosing the positive real axis of the $p_i^2$
plane, which starts at $\lambda^2-i\epsilon$ and ends at $\lambda^2+i\epsilon$,
but remains inside the overall contour $\gamma$ of Fig.~4.
Note that $\Gamma$ is not quite closed on the real axis.
We assume that $s$ and $t$ are large enough so that we may choose
\beq
\lambda^2\gg M_i^2\gg \mu^2,
\eeq
with $\mu$ the scale of power corrections, $\lambda^2$ given by
eq.~(\ref{lambda}) and $M_i$ the arguments of the transform.
We then define our modified ``Borel" transform of $T_4$ by
\beqa
\Phi_4(M_1^2,M_2^2,s,t) & \equiv & {1 \over (2\pi i)^2} \int_\Gamma
{dp_1^2 \over M_1^2} \int_\Gamma {dp_2^2 \over M_2^2}\
e^{-p_1^2/M_1^2}\left (1 - e^{-(\lambda^2-p_1^2)/M_1^2} \right )  \nonumber \\
&\ & \hskip 0.4 true in
\times e^{-p_2^2/M_2^2}\left (1 - e^{-(\lambda^2-p_2^2)/M_2^2} \right )
\ T_4(p_1^2,p_2^2,s,t)  \nonumber \\
& = & {1 \over (2\pi)^2}
 \int_{0}^{\lambda^2}{ds_1\over M_1^2}\int_{0}^{\lambda^2}{ds_2\over
M_2^2}\Delta_4(s_1,s_2,s,t)e^{-s_1/M_1^2 - s_2/M_2^2} \nonumber \\
&\ &
\hskip 0.4 true in  \times
\left (1 - e^{-(\lambda^2-s_1)/M_1^2} \right )
\left (1 - e^{-(\lambda^2-s_2)/M_2^2} \right ).
\label{phi4}
\eeqa
In the second equality,
the evaluation of the $p_i^2$ integrals has been carried out,
using Cauchy's theorem
and the representation eq.~(\ref{t4}) for $T_4$,
by simply shrinking the $\Gamma$ contours
to the fixed points $\lambda^2\pm i\epsilon$.  For $s_i$ real and
$\le \lambda^2$, we pick up a pole contribution, just as in eq.~(\ref{phi3})
for the three-point function, while for $s_i$ on the circular part
of the $\gamma$ contour in Fig.~4, we get no contribution, as the
$\Gamma$ contour shrinks to the infinitesimal segment between
$\pm i\epsilon$ without encountering any singularities.  At the
special points, $s_i=\lambda^2$, the factor $1-e^{-(\lambda^2-p_i^2)/M_i^2}$
vanishes at the $p^2_i=\lambda^2$ pole, so that the transform gets a
vanishing contribution from these points as well.
As usual,
$\Delta_4$ is the double discontinuity of $T_4$.
{}From eq.~(\ref{phi4}), we will
derive sum rules exactly analogous to the form factor result,
eq.~(\ref{3sr}), by equating the perturbative plus power corrections expression
for
$T_4 (s_1,s_2,s,t)$ with a phenomenological
parameterization of $\Delta_4$ in eq.~(19).  Let us see how this
works for the case at hand, the Compton scattering of a pion.

\section{\bf The Pion Compton Amplitude}

The pion matrix element
\beq
M_{\nu\lambda}= i\int d^4y e^{-iq_1 y} \langle
p_2|T\left(J_{\nu}(y)J_{\la}(0)\right)|p_1\rangle
\label{mnula}
\eeq
is the
object
we will study in terms of its invariant
amplitudes, for which we may construct sum rules as in the
previous section.
$M_{\nu\lambda}$ can be expressed in terms of two invariant amplitudes.
We use helicity form factors $H_1$ and $H_2$ defined by
the relation \cite{Landau}
\beq
M^{\lambda\mu}= H_1 e^{(1)\lambda}e^{(1)\mu} +
H_2 e^{(2)\lambda}e^{(2)\mu},
\label{h1h2}
\eeq
with
\beq
e^{(1)\lambda}={N^{\lambda}\over \sqrt{-N^2}} \,\,\,\,\,\,\,\,
e^{(2)\lambda}={P^{\lambda}\over \sqrt{-P^2}},
\label{edef}
\eeq

\beq
N^{\lambda}=\epsilon^{\lambda\mu\nu\rho}P_{\mu}r_{\nu}R_{\rho}
\,\,\,\,\,\,\,\,
P^{\lambda}=p_1^\lambda+ \nu p_2^\lambda - R^\lambda {p_1\cdot R + \nu p_2\cdot
R\over
R^2} ,
\eeq
where
\beq
\nu = {p_1\cdot p_2 - s_1 \over p_1\cdot p_2 - s_2}
\eeq
and where
\beq
R=q_1+q_2 \ \ \ \ \ \ \ \ r=q_2 -q_1.
\eeq
The ``polarization" vectors $e^{(i)}$
satisfy
\beqa
e^{(i)} \cdot q_1 &=& e^{(i)}\cdot q_2 = 0, \nonumber \\
e^{(i)}\cdot e^{(j)} &=& -\delta_{ij}\ .
\label{eortho}
\eeqa
Note that these relations hold for all positive $s_1=p_1^2$ and $s_2=p_2^2$,
whether
or not they are equal. We also note that $e^{(1)}$ serves as the normal
to the scattering plane defined by the Compton process, while
$e^{(2)}$ is a unit spacelike vector
whose spatial projection is in the scattering plane, but which is orthogonal to
the $q_i$'s.
Other expansions are also possible, but we choose this one for
definiteness and for its simplicity.

Now $M_{\nu\lambda}$ appears as the residue at the double pion
pole at $p_1^2=p_2^2=m_\pi^2$ in  the four-point
Green function $\Gamma_{\nu\la\mu\si}$ introduced in eq.~(\ref{gamma4}).
To be more specific, let us define a double discontinuity
$\Delta_{\nu\la\mu\si}$ of $\Gamma_{\nu\lambda\mu\sigma}$ across
the $p_1^2$ and $p_2^2$ cuts by
\beq
\Delta_{\nu\la\mu\si}
\equiv
\Gamma_{++} - \Gamma_{+-} - \Gamma_{-+} +\Gamma_{--}\ ,
\eeq
where we define $\Gamma_{++}=\Gamma_{\nu\la\mu\si}(p_1^2+i\epsilon,p_2^2+
i\epsilon)$,
and so on.  The poles may be isolated from the $x_0 \to - \infty$,
$y_0 \to \infty$ limit of $\Gamma_{\nu\la\mu\si}$ in
eq.~(\ref{gamma4}), by inserting complete sets of states in the resulting
time ordering, as
 \beqa
\Gamma_{\nu\lambda\mu\sigma} &=&
{i\over (2\pi)^6} \int d^4x d^4y d^4z\, d^4{p'}_1 d^4{p'}_2\,
\theta(y_0) \theta(-x_0)\nonumber \\
&\ & \times
e^{-ip_1\cdot x-iq_1\cdot z+ip_2\cdot y}
 \nonumber \\
&\ &\times\delta({p'}_1^2 -m_\pi^2)
\delta({{p'}_2}^2 - m_\pi^2) \langle0|\eta_{\sigma}(y)|p'_2\rangle\nonumber \\
&\ &\times
\langle p'_2|T(J_{\nu}(z)J_{\lambda}(0))|p'_1\rangle
\langle p'_1|\eta_{\mu}^\dagger(x)|0\rangle \ + \ \Gamma^{cont},
\eeqa
where $\Gamma^{cont}$ includes
contributions from the higher states and other time orderings,
which do not contribute to the double pion pole.
As usual, we express the matrix
elements of the axial currents as
\beq
\langle 0|\eta_{\sigma}(y)|p'_2\rangle=i f_\pi {p'_2}_{\sigma}e^{-i{p'}_2\cdot
y
},
\eeq
\beq
\langle p'_1|\eta_{\mu}^\dagger(x)|0\rangle =-i f_{\pi}
{{p'}_1}_{\mu}e^{i{p'}_1 \cdot x}.
\eeq
This relates the single pion states $p'_1, p'_2$ to the vacuum at times
$\pm \infty$,
with $f_\pi$ the pion decay constant.  We then derive the relation
\beqa
\Delta_{\nu\la\mu\si}
&=&
f_\pi^2\, p_{1\mu}p_{2\sigma}\, (2\pi)^2\delta(p_1^2-m_\pi^2)
\delta(p_2^2-m_\pi^2)\nonumber \\
&\ &\hskip 0.4 true in \times\ M_{\nu\lambda}(p_1,p_2,q_1)+\Delta^{cont},
\eeqa
with $M_{\nu\lambda}$ given by eq.~(\ref{mnula}), and with $\Delta^{cont}$
the double discontinuity of $\Gamma^{cont}$.

We now relate the tensor $\Delta_{\nu\la\mu\si}$
to the invariant amplitudes $H_1$ and $H_2$ of eq.~(\ref{h1h2}).
To do so, we begin by defining two projections,
\beqa
\Gamma^{(12)}(p_1,p_2,q_1)_{\mu\sigma}
 &\equiv& -g^{\nu\lambda}\Gamma_{\nu\la\mu\si},\\
\Gamma^{(1)}(p_1,p_2,q_1)_{\mu\sigma}
 &\equiv&  {e^{(1)}}^\nu {e^{(1)}}^\lambda
\Gamma_{\nu\la\mu\si}.
\eeqa
When the pion poles in $\Gamma^{(12)}$ and $\Gamma^{(1)}$ are isolated,
their coefficients will be proportional to $H_1+H_2$ and $H_1$, respectively.
Thus, we want to
identify a set of tensor structures in $\Gamma^{(12)}$ and $\Gamma^{(1)}$
in which the pion poles may appear.

Note that $\Delta_{\nu\la\mu\si}$
may be expanded in terms of tensors made from $g^{\alpha\beta}$ and
its external vectors, $q_i$ and $p_i$.  Following Refs.~\cite{IS} and
\cite{NR1}, we choose to expand in a basis made from the independent
vectors $p_1+p_2$, $p_1-p_2$ and $q_1$.
(The tensors $e^{(i)}_\mu e^{(i)}_\sigma$
built from the vectors in eq.~(\ref{edef}) may be expanded in
this basis
as well, when we include the metric tensor.)
We then seek a vector $n$, which has the properties
\beq
(n\cdot q_1)=(n\cdot p_1) - (n \cdot p_2)=n^2=0.
\label{ndef}
\eeq
It is easy to verify that no such $n^\mu$ exists with real coefficients.
In fact we can
immediately give an explicit expression for $n^\mu$ in terms of the
polarization vectors $e^{(i)}$ above, as
\beq
n^\mu = \biggl (e^{(2)} \pm ie^{(1)} \biggr )^\mu\ .
\label{nconstruct}
\eeq
The specific values of the $n^\mu$ are frame-dependent of course,
and we need not give them here.
We note, however,
that either choice
of sign in eq.~(\ref{nconstruct})
will give the same (real) answers below.  The overall normalization
of $n^\mu$ also cancels
in the sum rule result.

If we saturate the $\mu$ and $\sigma$ indices
with an $n$ that satisfies eq.~(\ref{ndef}),
we will project onto the tensor structure $(p_1+p_2)_\mu(p_1+
p_2)_\sigma$ only, whose corresponding invariant functions we
denote as
\beqa
\Gamma^{(12)}_4(p_i^2,s,t) &\equiv& n^\mu n^\sigma\Gamma^{(12)}
(p_1,p_2,q_1)_{\mu\sigma}\ ,\\
\Gamma^{(1)}_4(p_i^2,s,t) &\equiv& n^\mu n^\sigma
\Gamma^{(1)}(p_1,p_2,q_1)_{\mu\sigma}\ .
\eeqa
Finally, taking as
above the double discontinuities, denoted as $\Delta_4^{(i)}$,
we isolate the desired invariant functions
\beqa
\Delta_4^{(12)}(p_i^2,s,t)
&=&
f_\pi^2\, (n\cdot p_{1})(n\cdot p_{2})\, (2\pi)^2\delta(p_1^2-m_\pi^2)
\delta(p_2^2-m_\pi^2) \nonumber \\
&\, & \times \left ( -g^{\nu\lambda}M_{\nu\lambda} \right )\, +
\Delta^{(12)con
t}_4
\nonumber \\
&=&
f_\pi^2\, (n\cdot p_{1})(n\cdot p_{2})\, (2\pi)^2\delta(p_1^2-m_\pi^2)
\delta(p_2^2-m_\pi^2) \nonumber \\
&\, & \times \left ( H_1(s,t)+
 H_2(s,t) \right )\, +  \Delta_4^{(12)cont},
 \eeqa
and similarly
\beqa
\Delta_4^{(1)}(p_i^2,s,t)
&=&
f_\pi^2\, (n\cdot p_{1} )(n\cdot p_{2})\, (2\pi)^2\delta(p_1^2-m_\pi^2)
\delta(p_2^2-m_\pi^2) \nonumber \\
&\, & \times\  H_1(s,t)\, +  \Delta_4^{(1)cont}.
 \eeqa

The sum rules for the $\Gamma_4^{(i)}$ are derived as for $\Gamma_3$
above.  Consider, for instance, $\Gamma^{(12)}_4$.
First, we approximate $\Gamma_4^{(12)}$ by perturbation theory plus
power corrections for a choice of contour $\gamma$ in eq.~(\ref{t4}) that stays
away from the real axis, except at $p_i^2=\lambda^2$,
\beq
\Gamma_4^{(12)}(p_1^2,p_2^2,s,t)=\Gamma_4^{(12)pert}(p_1^2,p_2^2,s,t)
+ {\rm power\ corrections}.
\eeq
This is analogous to eq.~(\ref{Cb}) for the three point function.
Next, choosing $\gamma$ in eq.~(\ref{t4}) on the real axis, we approximate
its discontinuity  by a
``first resonance plus continuum''
form, analogous to eq.~(\ref{cont}):
\beqa
\Delta_4^{(12)} &=&
f_\pi^2\, (n\cdot p_{1})(n\cdot p_{2}) (2\pi)^2\delta(p_1^2-m_\pi^2)
\delta(p_2^2-m_\pi^2) \nonumber \\
&\ &\ \times \left ( H_1(s,t)+
 H_2(s,t) \right ) \nonumber \\
 &\, &\ \  +\ \Delta_4^{(12)pert}\left ( 1 - \theta(s_0-s_1)
 \theta(s_0-s_2) \right ).
 \eeqa
Combining these expressions in the transform of eq.~(\ref{phi4}), we derive the
full sum rule as
\beqa
&&f_\pi^2\, (n\cdot p_{1})(n\cdot p_{2})
\left ( H_1(s,t)+
 H_2(s,t) \right ) \nonumber \\
&&\quad  = \int_{0}^{s_o}ds_1\int_{0}^{s_o}ds_2\ \Delta^{(12)pert}_{4}
(s_1,s_2,s,t)e^{-s_1/M_1^2 -s_2/M_2^2} \nonumber \\
&& \hskip 0.4 true in \times
\left (1 - e^{-(\lambda^2-s_1)/M_1^2} \right ) \left (1 -
e^{-(\lambda^2-s_2)/M_
2^2} \right )
\nonumber \\
&& \quad +
f_1(s,t,M_i^2) {\langle 0|\alpha_s
G_{\mu\nu}G^{\mu\nu}|0 \rangle \over (M_1^2)^2} + f_2(s,t,M_i^2){\langle
0|\alpha_s (\bar{q}q)^2|0 \rangle \over (M_1^2)^3} +...,
\label{compsr}
\eeqa
where we have written out the power corrections
schematically,
as in eq.~(\ref{conden}). The functions $f_i(s,t,M_i^2)$ may be
calculated as for the three-point function \cite{IS,NR1}.  We reserve
this calculation for future investigation.

As in the form factor case of Section 2 above, $s_0$ should be
fixed by demanding best stability with respect to variations in
$M^2=M_1^2=M_2^2$.
We shall assume that the power corrections are small enough
so that, as we increase the $M_i^2$, still keeping $s/M_i^2$
large, there is a range where they
are negligible compared to the perturbative contributions on the
right-hand side of eq.~(\ref{compsr}).  Then we may
write a ``local duality" sum rule, without power corrections:
\beq
f_\pi^2\, (n\cdot p_1) (n\cdot p_2)\,
\left ( H_1(s,t) +
 H_2(s,t) \right )
= \int_{0}^{s_o}ds_1\int_{0}^{s_o}ds_2\Delta^{(12)pert}_{4}
(s_1,s_2,s,t)\ .
\label{ldsr}
\eeq
We should emphasize that the derivation of this
result from eq.~(\ref{compsr}) involves assumptions on the
behavior of the condensates.
The right hand side of eq.~(\ref{ldsr}) is approximated only by
the local duality integral, and is to be calculated using
perturbation theory. This is the subject of the following section.

\section{\bf Local Duality Sum Rules and Asymptotic Behavior}

The lowest order perturbative spectral weights $\Delta_4^{(i)\rm pert}$ can
be computed from Figs.~5(a) and 5(b) by use of Cutkosky rules.
The remaining lowest order diagrams, in which $q_2$ and $p_2$ are exchanged,
do not contribute to the double discontinuity in this kinematic region.
The double discontinuity
in $p_1^2$ and $p_2^2$ is found by simply replacing the propagators
that carry momenta $k^\mu$, $(p_1-k)^\mu$ and $(p_2-k)^\mu$ by
factors $(2\pi)\delta_+(k^2)$, and so on, where the plus indicates
that energy flows in the same direction as for $p_1$ and $p_2$.
The procedure is similar to the one adopted in
the case of the triangle diagram \cite{IS,Eden} for 3-point
functions.

We then write $\Delta_4^{(i)pert}$ as the sum of the two spectral functions
$\Delta_a^{(i)pert}$ and $\Delta_b^{(i)pert}$ found from the diagrams of
Fig.~5,
with an overall constant separated out,
\beqa
&&\Delta_4^{(i)pert}(s_1,s_2,s,t) \nonumber \\
&& \quad \quad ={5\over 3 (2\pi)^3}
(\Delta_a^{(i)pert}(s_1,s_2,s,t) +
\Delta_b^{(i)pert}(s_1,s_2,s,t)).
\label{Deltadef}
\eeqa
The extra factors are from the flavor structure of the current
and the color traces, as well as from the factors of $(2\pi)^{-1}$
associated with the loop integral.  A factor of
$(2\pi)^{-2}$ associated with the dispersion relation
eq.~(\ref{t4}) has also been absorbed into the
definition of the spectral function.

For specific calculation, we found it useful to work in a ``brick-wall" frame
for $p_1$ and $p_2$, with $p_{1\perp}=
p_{2\perp}=0$; the corresponding kinematics is described
in some detail in Appendix B.
Let us consider the $-g^{\nu\lambda}$ projection, $\Delta_4^{(12)}$,
related by the sum rule to the combination $H_1+H_2$.
We have, in the normalization of eq.~(\ref{Deltadef}),
\beqa
\Delta^{(12)pert}_a&=& 16 \int d^4 k\ { \delta_+(k^2)\delta_+((p_1-k)^2)
\delta_+((p_2-k)^2)\over (p_1 - k + q_1)^2}\nonumber \\
&\ &\quad  \times\, (n\cdot k)\, [(q_1\cdot p_1 +q_1 \cdot p_2)\,
(n\cdot p_1 - n \cdot k)
\nonumber \\
&\ &\quad \quad  - (q_1\cdot k) (n\cdot p_1+n\cdot p_2 - 2n\cdot k) ],
\label{12delt}
\eeqa
with a related form for $\Delta^{(12)pert}_b$, found
by replacing $q_1$ by $-q_2$, and hence $s$ by $u$ in the
resulting answers. For $\Delta^{(1)}_a$, we have
\beqa
\Delta^{(1)pert}_a &=& {1\over 2} \Delta^{(12)pert}_a \nonumber \\
&\ & + 8 \int d^4 k\, { \delta_+(k^2)\delta_+((p_1-k)^2)
\delta_+((p_2-k)^2)\over (p_1 - k + q_1)^2}\nonumber \\
&\ &\quad  \times\, (e^{(1)}\cdot k)\, (n\cdot k)\,
 \biggl \{ (e^{(1)}\cdot k)\, \left ( n\cdot p_1+n \cdot p_2 - 2k\cdot n
\right )  \nonumber \\
&\ &\quad \quad \quad \quad + (e^{(1)}\cdot n)\, \left ((p_1-k)\cdot
(p_2-k)\right ) \biggr \},
\label{1delt}
\eeqa
where we note that the second
term is at least quadratic in $k$.
The local duality sum rules are  found by substituting the results of these
integrals into eq.~(\ref{ldsr}).  For instance, we find
\beq
{f_\pi}^2(H_1 + H_2)=\int_{0}^{s_0}ds_1 \int_{0}^{s_0}ds_2\
\rho^{pert}_4(s_1,s_2,s,t),
\label{h12ldsr}
\eeq
where
\beq
\rho^{pert}_4(s_i,s,t) = {1 \over (n\cdot p_1)(n\cdot p_2)}
\Delta_4^{(12)pert}(s_i,s,t).
\label{rho4pert}
\eeq

The integrals in eqs.~(\ref{12delt}) and (\ref{1delt}) are all
elementary.  The specific complete integrals and results
are given in Appendix C.  Here we shall only discuss the leading
behavior at fixed angles.  It is easy to show that it comes
from the term linear in $n\cdot k$, and is specified by
\beqa
\rho^{(12)pert}_4
&\approx& {10\over 3(2\pi)^2}\left ({s_1 +s_2 \over (-t)^2} \right )
 \left ( {(s-u)^2\over 2s(-u)} \right ) \nonumber \\
&\approx& 2\rho^{(1)pert}_4.
\label{rhohien}
\eeqa
Here we have used eq.~(\ref{rho4pert}), which defines $\rho$ in terms
of $\Delta$, and have combined the contributions from both diagrams of
Fig.~5.
This result exhibits a comparable power suppression to the spectral
weight encountered in the form
factor \cite{IS,NR1}. (Although the
three point function involves
one less external photon, a factor of $(p_1+p_2)^\lambda$
is factored from its amplitude, so that the remaining
form factor has the same scaling behavior as in Compton
scattering.)
To be specific, the lowest-order perturbative spectral weight
in the latter case may be written as exactly \cite{NR1}
\beqa
&&{\rho_\pi}^{pert}_3(s_1,s_2,t) \nonumber \\
&& \quad \quad ={3\over 2 \pi^2} t^2 \biggl( \left ({d\over dt}\right )^2 +
{t\over 3} \left ({d\over dt}\right )^3 \biggr)
{1\over ((s_1 + s_2 - t)^2 - 4 s_1 s_2)^{1/2}}.
\eeqa
Asymptotically at large $t$ this function
has the characteristic behavior
\beq
\rho_\pi = {1\over \pi^2}
\biggl ({3(s_1+s_2)\over  t^2} + O(1/t^3)
\biggr ).
\eeq
This is suppressed in $t$ by one additional
power
compared to the prediction derived from the dimensional counting
rules \cite{Matveev,Farrar,pQCDff}.

The leading Compton scattering integral in eq.~(\ref{h12ldsr}) can be evaluated
trivially in the local duality approximation, in which we neglect
$s_i/M_i^2$ compared to unity.
We get, for $H_1+H_2$, for instance,
\beqa
(H_1 + H_2){f_\pi}^2 & = &
 {10\over 3(2\pi)^2}
\int_{0}^{s_0}ds_1\int_{0}^{s_0}ds_2 {(s_1+s_2) \over (-t)^2 }
\biggl ( 1 - {s^2+u^2 \over 2su} \biggr )
\nonumber \\
&=&  {10\over 3(2\pi)^2} \biggl ( {s_0^3 \over (-t)^2} \biggr )
\biggl ( 1 - {s^2+u^2 \over 2su} \biggr )\ .
\label{hexpansion}
\eeqa
As in the case of
the form factor, the result is suppressed by an additional  power of $t$
compared to the ($1/t$) perturbative behavior expected from the
dimensional counting rules.
This may be understood in terms of the result that the
true asymptotic behavior of form factors is dominated
by hard gluon exchange \cite{pQCDff}.  It is easy to see (Appendix C)
that as $s\rightarrow \infty$, with $s_1$ and $s_2$
fixed, the momentum of the unscattered
parton ($k$ in Fig.~5) is forced to zero,
while the incoming and
outgoing momenta of the scattered parton approach the
momenta of the external ``pions".  This
description of elastic scattering, mediated by
a single parton
that carries essentially all the energy of a hadron, is often
called the ``Feynman" mechanism \cite{Fey}.  Although it
cannot give the true asymptotic behavior, it may be important
at intermediate energies \cite{Radyconf}.

\section{\bf Conclusions}

In this paper, we have shown that QCD sum rules
can be formulated for pion
Compton scattering, generalizing previous results on meson form
factors.  We have shown that it is possible to
write down a dispersion
representation of the scattering amplitude on a finite contour in the
complex plane of the external masses, and we have explicitly
calculated the leading behavior of the spectral function.
To derive a sum rule, we found it useful to introduce a modified definition
of the Borel transform through an integral representation, eq.~(\ref{phi4}).
The sum rule, eq.~(\ref{compsr}), in the local duality limit, shows suppression
by an additional power of $t$ compared to the perturbative result,
as shown in eq.~(\ref{hexpansion}).

In the future, we hope to pursue the QCD sum rule procedure
futher, with the inclusion of power corrections assoicated
with  the condenstates that appear in  power corrections from
the operator product expansion.  Here, however, we have demonstrated that
such a program is possible.

The approach that we have introduced in this paper
can also be generalized to the more interesting case of
proton Compton scattering \cite{CC}, where experimental data are
availiable, while a direct comparison with perturbative QCD
results based on leading power factorization theorems \cite {KNizic}
is possible. Although the leading power result will, by definition,
dominate for sufficiently large $t$, we expect the sum rule result to be a
valuable tool in the interpretation of realistic experiments. Of particular
 interest will be the angular
 dependence of the sum rule result, which may
 distinguish between leading-power factorization theorems
 and the
 ``Feynman mechanism" that dominates the sum rule
spectral functions.  This could help us to understand the
roles of these two formalisms for describing hadronic
elastic scattering at available energies.

{\it Acknowledgments.}
 This work was supported in part by the National Science Foundation, grants
PHY-9211367 and
91080541, by
the U.S. Department of Energy under contract DE-AC05-84ER40150, and
and by the Texas National Research Laboratory.

\appendix

\section{Analyticity and Infrared Safety}

To derive a sum rule for the Compton scattering amplitude, we must be able
to prove, first, that the amplitude satisfies the dispersion relation,
eq.~(\ref{t4}).
In addition, we must check that for nonzero values of $p_i^2$ the
amplitude is infrared safe; i.e. that its perturbative expansion is free
of infrared divergences.
Otherwise, the lowest-order term in the expansion will be
swamped by divergent higher-order corrections.
The demonstrations of these two features of the fixed-angle Compton
amplitude are related, and we present them together.  We will, of
course, work in perturbation theory.

Eq.~(\ref{t4}) is consistent with singlularities at positive values of $p_1^2$
and/or $p_2^2$, corresponding to normal and anomalous thresholds.  Let us
recall \cite {Eden} that the condition for a singularity in a Feynman amplitude
is that the momentum integrals be ``pinched" between coalescing poles
from the propagators of internal lines.  If internal lines are not on-shell
the integrand is finite; if the momentum contours are not pinched, we may
deform the momentum integrals into another region where the integrands are
finite.

A useful way of characterizing points in momentum space where some set
of lines go on-shell is through ``reduced diagrams", in which all lines
that are off-shell are shrunk to points. Now suppose the resulting
reduced diagram represents a point in momentum space at which the
contours are truly pinched.  Then a theorem due to Coleman and Norton
\cite {CN} states that
the diagram corresponds to a classical scattering process, in which all
vertices correspond to points in space-time, connected by the motion
of on-shell, free, classical particles, whose velocity is determined by their
mass and momentum.  This theorem turns out to be a very useful tool in the
analysis that follows.

Let us start by considering Compton scattering with space-like  photons,
$q_i^2 < 0$.  We will relax this condition later.  Now suppose $p_1^2$ is
either negative, or complex, but with an amplitude that
obeys the restriction,
\beq
|p_i^2| \le \lambda^2\ ,
\label{inequal}
\eeq
with $\lambda^2=(s+t)/4$, as in eq.~(15), so that
\beq
Re(u) < 0.
\eeq
We will be trying to show that for such a $p_1^2$ the
Compton amplitude has cut plane analyticity in the $p_2^2$
plane when $p_2^2$ also satisfies eq.~(\ref{inequal}).

Let us see what kind of
physical-scattering reduced diagrams, which
potentially give singularities,
can include external currents with
momenta in the ranges just described, all the time keeping $s$ and
$t$ fixed and large.  Because $s$ is positive and $t$ negative, we
may assume that the energies of $p_1 \dots q_2$ remain positive.  The
first vertex in the reduced diagram must therefore result from the
scattering of $p_1$ and $q_1$,
or of some subset of particles from one of their virtual states.  When
$q_1^2<0$ and $p_1^2<0$ or complex, however, neither the photon nor
$p_1$ can decay into
only on-shell physical particles as part of the physical
process. Since $t<0$, $p_1$ also cannot decay by emitting the (possibly
unphysical) momentum $p_2$, and because $u<0$ (in the region we are
considering), it cannot decay into a physical state by emitting
$q_2$, even if $p_1^2$ is real.

We may now take the limit $q_1^2\rightarrow 0$, which brings us up to, but not
 beyond, thresholds in the $q_1^2$ channel.  The virtual states into which the
 incoming photon can ``decay" consist of collinear massless particles,
 which propagate together at the speed of light.  In any space-time
 picture, they therefore act in the same manner as a single particle.
The first vertex in the reduced diagram must therefore
involve the direct scattering of $q_1$
(or its collinear, lightlike decay products) and $p_1$ at a point, and $all$
the remaining particles must emerge from that point.  Any situation
such as this, in which all particles emerge from a single point in
space-time, strongly reduces the number of possible reduced diagrams
\cite {St77b}. In fact, in this case, the only possible reduced
diagrams corresponding
to singular points involve no further ``hard" scatterings with nonzero
momentum transfer.  All further interactions involve the local interactions
of massive particles, relatively at rest,  of exactly collinear-moving massless
particles, and/or the emission and absorption of exactly zero-momentum
particles.  In our case, this means that we may only
 have thresholds when $p_2^2
=(\sum_im_i)^2$, for some set ${m_i}$ of the masses of virtual
particles.  Similarly, for $q_2^2\rightarrow 0$ we may have on-shell massless
virtual particles moving exactly parallel to $q_2$.
We need not consider soft particles that connect lines moving parallel
to $p_2$ with those moving parallel to $q_2$, because, since
the photon is color and charge neutral, its couplings to soft
particles vanish on-shell \cite {St77a}.  In addition, since we keep
$q_2^2$ fixed, we only need to consider the singularities associated
with $p_2$, that is, its normal thresholds for $p_2^2\ge 0$.  Clearly,
analogous reasoning, working backwards
from the final toward the initial state can be used to show that
the amplitude has cut-plane analyticity in $p_1^2$ for $p_2^2$
fixed and off the real axis. This is the first result we set out to prove.

Given the preceeding arguments, the proof of infrared safety is not
difficult.  We are interested in the behavior of the amplitude around the
contours at $|p_i^2|=\lambda^2$.  In this region, we want to work in
perturbation theory with massless internal lines.  It is easy to see that
the only reduced diagrams which correspond to points at which the momentum
contours are pinched consist of a single hard scattering vertex, and of
two ``jet" subdiagrams, connected to the external photons, and each consisting
entirely of lines collinear to that photon.  There may also be
zero momentum lines connecting to two jet subdiagrams and/or the hard
vertex.  Note that because we work to lowest order in QED, the jet
subdiagrams are not photon self-energies, but must attach to the hard
scattering by at
least two finite-momentum lines each.  The
second result claimed above, the
infrared finiteness of such an amplitude, then follows directly from the
arguments given in \cite{St77a} for $e^+e^-$ annihilation into photons.

\section{Kinematics}

In this appendix we discuss the kinematics of the pion Compton
scattering process in Fig.~3.

Let $q_1$ and $q_2$ be the momenta of the incoming and outgoing photon,
respectively, which are assumed to be on-shell ($q_1^2=q_2^2=0$).
Let also $p_1$ and $p_2$ be the momenta of the incoming and outgoing pion.
The external pion states are off-shell and are characterized by the
invariants $s_1=p_1^2$ and $s_2=p_2^2$.

We define as usual the Mandelstam invariants
\beq
s=(p_1+q_1)^2, \,\,\,\,\,\,\,\,\,\, t=(q_2-q_1)^2,
\,\,\,\,\,\,\,\,\,\,\,
u=(p_2-q_1)^2,
\eeq
with the usual relation
\beq
s+t+u=s_1+s_2.
\eeq
We consider both $s$ and $t$ to be very large and in the physical region
$s>0,t<0$.

 It is also convenient to introduce
light cone variables for the pion and photon momenta as follows:
\beqa
p_1&=&p_1^+\bar{v} + p_1^- v, \quad \quad \quad  p_2=p_2^+\bar{v} +p_2^- v,
\\
q_1 &=& q_1^+ \bar{v} +q_1^- v +q_{1\perp},  \\ \nonumber
 \bar{v}&=&{1\over\sqrt{2}} (1,1,{\bf 0}_\perp), \quad \quad \quad
 v={1\over\sqrt{2}} (1,-1,{\bf 0}_\perp), \nonumber \\
 &\ & \hskip 0.75 true in q_{1\perp}\cdot n^{\pm} = 0.
\eeqa
In the definitions of $v$ and $\bar{v}$, the order of
components is $(0,3,2,1)$.
The momenta of the pions are, in this frame, purely longitudinal.
We can choose a particular brick-wall frame in which $p_1^+=p_2^-=Q$.
Then we have
\beq
p_1=Q \bar{v} +{s_1\over 2 Q} v, \,\,\,\,\,\,\,\,\,\,\, p_2={s_2\over 2
Q} \bar{v} + Q v.
\eeq

A covariant expression for the momentum scale $Q$ can be found from the
relation
\beq
t=(p_2-p_1)^2=s_1+s_2 -2 Q^2- {s_1 s_2 \over 2 Q^2}. \eeq
We easily find, by solving for $Q^2$,
\beq
Q^2={1\over 4} (s_1+s_2-t + \delta) = {1 \over 4} (s+u+\delta)\ ,
\eeq
where we have defined
\beq
\delta=\sqrt{(s_1+s_2-t)^2 -4 s_1 s_2} = 2Q^2 - {s_1s_2 \over 2Q^2}.
\label{deltadef}
\eeq
In the scattering process at high energy and
fixed angle, $Q^2$ is a large
parameter.  Note that in this notation $t=-2Q^2$ only when $s_1=s_2=0$.
 In this frame we also easily get
\beq
u=(p_2-q_1)^2=2 Q^2-s +{{s_1 s_2}\over 2 Q^2}.
\eeq

Covariant expressions for $q_1^\pm$ and $q_2^\pm$ can also be found
in the form
\beq
q_1^+={(s-2 Q^2)(2 Q^2-s_2)\over 2Q\delta},
\eeq

\beq
q_1^-={(2 Q^2-s_1)(2 Q^2 s-s_1 s_2)\over 4Q^3\delta}.
\eeq
The magnitude of the transverse component $q_{1\perp}$ of the momentum of
the incoming photon is then given by
\beq
q_{1\perp}\equiv q_\perp ={\sqrt{(-t)(s-2 Q^2)(2 Q^2 s-s_1 s_
2)}\over
\sqrt{2}Q\delta }.
\eeq
Similar formulas hold for the momentum of the outgoing photon:
\beqa
 q_2^+&=&{(2 Q^2-s_2)(2 Q^2 s-s_1 s_2)\over 4Q^3\delta},\nonumber \\
q_2^-&= &{(s-2 Q^2)(2 Q^2-s_1)\over 2Q\delta}.
\eeqa

\section{Computation of the Spectral Weights}

In this appendix, we discuss the evaluation of the integrals
in eq.~(\ref{12delt}) and their asymptotic behavior at high energy.
The results for the crossed diagram are carried out analogously.

We first note that we may use the identity
\beq
2q_1\cdot k = s-s_1 - (p_1-k+q_1 )^2
\label{aident}
\eeq
to rewrite eq.~(\ref{12delt}) in terms of integrals
for the fully on-shell box and triangle diagrams that are no
more than quadratic in $k^\mu$ in the numerator.  Thus, we have
\beqa
\Delta^{(12)pert}_a
&=& 16(q_1\cdot p_1+q_1 \cdot p_2) \int d^4k\ I_s^{(4a)}\ \left (\,
(n\cdot k)(n\cdot p_1) - (n\cdot k)^2\, \right )
\nonumber \\
&\ & + 8 \int d^4k\ [\, I_s^{(3)} - (s-s_1)I_s^{(4a)}\, ]\nonumber \\
&\ & \quad \quad \times \left ((n\cdot k) (n\cdot p_1+n\cdot p_2) -
2(n\cdot k)^2\, \right ),
\label{quad12}
\eeqa
where $I^{(3)}$ and $I_s^{(4a)}$ are the integrands for the
double discontinuities of the triangle and box diagrams, respectively,
\beqa
I_s^{(3)} &=& \delta_+
(k^2)\delta_+((p_1-k)^2)\delta_+((p_2-k)^2)\ ,
\nonumber \\
I_s^{(4a)} &=& {\delta_+
(k^2)\delta_+((p_1-k)^2)\delta_+((p_2-k)^2)\over
(p_1-k + q_1)^2}.
\label{Idef}
\eeqa
The relevant integrals are now
\beqa
v^\mu &=& \int I^{(3)}_s k^\mu \nonumber \\
V^{\mu\nu} &=& \int I^{(3)}_s k^\mu k^\nu
\label{veedef}
\eeqa
for the triangle  and
\beqa
w_a^\mu &=& \int I^{(4a)}_s k^\mu \nonumber \\
W_a^{\mu\nu} &=& \int I^{(4a)}_s k^\mu k^\nu
\label{weedef}
\eeqa
for the box.

Before discussing the foregoing integrals, it is
useful to describe the corresponding scalar integrals for
the scalar triangle and the scalar
box.  For the triangle we have
\beq
\Delta_{3}=\int d^4 k\, \delta_+(k^2)\delta_+((p_1-k)^2)
\delta_+((p_2-k)^2)={\pi\over 2 \delta},
\eeq
where $\delta$ has been defined in eq.~(\ref{deltadef}).
The integral for $\Delta_3$ can be simply evaluated by going to the
brick-wall frame defined above; $2\delta$ appears as the Jacobian of
the transformation from the
components $k^\mu$ to the variables $k^2, p_1\cdot k$, and $p_2\cdot k$,
leaving one trivial angular integral,
\beq
\Delta_{3}={1\over 4 \delta}\ \int_0^{2\pi} d\phi\ ,
\eeq
with $\phi$ the angle in transverse space.
Interestingly, the discontinuity on the cut is given
by the same result for massive as for massless quarks.

For the scalar box diagrams in Figs.\ 5(a) and 5(b), three
internal lines
are on shell, while the
upper internal line is off-shell,
\beq
\Delta_a^{sc}\ (s_1,s_2,s,t)=\int d^4 k\, {\delta_+
(k^2)\delta_+((p_1-k)^2)\delta_+((p_2-k)^2)\over
(p_1-k + q_1)^2}.
\eeq
Three angular integrals are trivially done using the change of variables
mentioned above, but the remaining angular integral is now nontrivial,
\beq
\Delta_a^{sc} (s_1,s_2,s,t)={1\over 4 \delta}\ \int_0^{2\pi} d\phi
\biggl ( { 1\over s-s_1 -2(q_{1L}\cdot k_L - q_\perp k_\perp cos\phi )} \biggr
)
\ ,
\eeq
where we define
\beq
q_{1L}\cdot k_L \equiv q_1^+k^-+q_1^-k^+\ ,
\eeq
which is a number that is fixed in the brick-wall frame according to
the values given in Appendix B and the
mass shell delta functions (see below).
The angular integral is a standard one,
\beq
\int_0^{2\pi} d\phi\, \biggl ( { 1 \over A + Bcos\phi} \biggr ) =
{2\pi \over (A^2-B^2)^{1/2}}\ .
\label{angint}
\eeq
For the particular constants
\beqa
A&=&s-s_1-2q_{1L}\cdot k_L= {(-t) \over \delta^2}
[s(s+u)-2s_1s_2]\ ,
\nonumber \\
B&=&2q_\perp k_\perp={2(-t) \over \delta^2}[s(-u)
+s_1s_2]^{1/2}[s_1s_2]^{1/2}\ ,
\eeqa
we find the following, rather unobvious, identity,
\beq
{1 \over (A^2-B^2)^{1/2}}=
{1 \over ([s-s_1-2q_{1L}\cdot k_L]^2-[2q_\perp k_\perp]^2)^{1/2}}=
{\delta \over s(-t)}\ .
\eeq
Thus, at lowest order, the box and crossed box
 are given by the remarkably simple forms,
\beq
\Delta_a= - {\pi\over 2 s t}\ ,
\eeq
\beq
\Delta_b=-{\pi\over 2 u t}\ ,
\eeq
where we have used the crossing relation between the two.

In deriving the spectral weight for the box diagram with
fermions, it is useful to know expressions for the components of
$k^\mu$ in the brick-wall frame kinematics of Appendix B,
\[
k^+={s_2(Q-s_1/2Q)\over \delta},
\]
\beq
k^-={s_1(Q-s_2/2Q)\over \delta},
\eeq
as fixed by the delta functions in the integral.
We notice that both components are small and $O(1/Q)$.
Similarly, we have the simple expression
\beq
{k_\perp}^2
 = {s_1s_2(-t) \over \delta^2}.
\eeq
Note that all the components of $k^\mu$ vanish in the limit $Q\rightarrow
\infty$, as is appropriate for the Feynman mechanism (Section 5).

The integrals that define the spectral functions, eq.~(\ref{12delt}),
are easily reduced
in the brick-wall frame to the factor $1/4\delta$ times
angular integrals which generalize eq.~(\ref{angint}),
\beq
I_a=\int_0^{2\pi} d\phi\, \biggl ( { cos^a\phi \over A + Bcos\phi} \biggr )\ ,
\eeq
with $a=0,1,2$ and with $A$ and $B$ given as above.  For completeness,
we list the elementary results,
\beqa
I_0 &=& {2\pi \over (A^2-B^2)^{1/2}}={2\pi\delta\over s(-t)}\ ,\nonumber \\
I_1 &=& { 2\pi A\over B(A^2-B^2)^{1/2}} - {2\pi \over B}=
{4\pi\delta \over s^2(-t)}{\sqrt{s(-u)s_1s_2+s_1^2s_2^2}
\over 4Q^2 -2s_1s_2/s}\ ,\nonumber\\
I_2 &=&  { 2\pi A^2\over B^2(A^2-B^2)^{1/2}} - {2\pi A\over B^2}=
{2\pi\delta \over s^2(-t)}{s(s+u)-2s_1s_2 \over
4Q^2-2s_1s_2/s}\ .\nonumber\\
\eeqa
We note that in the frame we work, $A\gg B$, and that while $I_0$ and
$I_2$ behave as $1/A$ for large $A$, $I_1$ goes as $1/A^2$.
In terms of these expressions, we find,
\beq
v^\mu = {\pi \over \delta^3} [\, s_2(p_1\cdot p_2-s_1)p_1^\mu
+ s_1(p_1\cdot p_2-s_2)p_2^\mu\, ]\ ,
\eeq
\beq
V^{\mu\nu} =
{2\delta \over \pi}\biggl ( v^\mu v^\nu - {1\over 2}d^{\mu\nu}v^2
\biggr )\, ,
\eeq
where $d^{11}=d^{22}=1$, with all other components zero, and
\beq
w^\mu={\delta \over s(-t)}v^\mu+{1\over 2\delta}
\sqrt{{s_1s_2 \over s(-u)+s_1s_2}}
q_\perp^\mu\ ,
\eeq
\beqa
W^{\mu\nu}&=&{2s(-t) \over \pi}
\bigl ( w^\mu w^\nu - \delta_{\mu 2}\delta_{\nu 2}(w_2)^2 \bigr )
\nonumber \\
&\ & \quad - {s_1s_2(-t) \over 4\delta^3}(I_2\delta_{\mu 2}\delta_{\nu 2}
+(I_0-I_2)\delta_{\mu 1}\delta_{\nu 1})\ .
\eeqa

The leading high-energy behavior of
$\Delta_a^{(12)pert}$ in eq.~(\ref{12delt}) comes entirely
from the term
\beqa
\Delta_a^{(12)pert} &=& 8(s-u)(n\cdot w_a)\, (n\cdot p_1) + \cdots \nonumber \\
&\sim& 4(s-u) \left ({\pi \over s(-t)^2}(s_1+s_2) \right )
(n\cdot p_1)(n\cdot p_2)\ ,
\eeqa
where we have used $(n\cdot p_1)=(n\cdot p_2)$.  Note that by
eq.~(\ref{rho4pert}), the factor $(n\cdot p_1)(n\cdot p_2)$
cancels in the sum rule.
Similarly, we find for the crossed
diagram $\Delta_b^{(12)pert}$:
\beq
\Delta_b^{(12)pert}
\sim 4(u-s)\left ( {\pi \over u(-t)^2}(s_1+s_2) \right )
(n\cdot p_1)(n\cdot p_2)\ .
\eeq
Finally, using eqs.~(\ref{Deltadef}) and
(\ref{rho4pert}), we get the asymptotic spectral function
quoted in eq.~(\ref{rhohien}).

\newpage


\newpage
\noindent
{\bf Figure Captions}
\bigskip

\noindent
1. The integration
contours for the Borel transform in the $p_i^2$ plane,
eq.~(\ref{borel}).
\smallskip

\noindent
2. Lowest order perturbative contribution to the pion form factor.
\smallskip

\noindent
3. The pion Compton scattering amplitude.
\smallskip

\noindent
4. The integration contours for the
Compton scattering amplitude.  Here $\gamma$ is the dispersion
contour of eq.~(\ref{dispers}), and $\Gamma$ the
transform contour of eq.~(\ref{phi4}).
\smallskip

\noindent
5.  Lowest order contributions to the spectral weight for Compton
scattering.


\newpage
\begin {thebibliography}{99}

\bibitem{SVZ} M.A.\ Shifman, A. I. Vainshtein and V.I. Zakharov, Nucl.
Phys. B147 (1979) 385, 448, 519.

\bibitem{IS} B.L.\ Ioffe and A.V. Smilga, Phys. Lett. 114B (1982)
353; Nucl. Phys. B216 (1983) 373.

\bibitem{NR1}V.A.\ Nesterenko and A.V. Radyushkin, Phys. Lett. 115B
(1982) 410; JETP Lett. 35 (1982) 488.

\bibitem{BI} V.M.\ Belyaev and B.L.\ Ioffe, Nucl.\ Phys.\ {B310}
(1988) 548.

\bibitem{ES} L.L.\ Enkovskii and B. V. Struminskii,
Theo.\ and Math.\ Phys.\ 57 (1983) 979 (Teor. Mat. Fiz. 57
(1983) 41).

\bibitem{LST} A.A.\ Logunov, L. D. Soloviev and A. N. Tavkelidze, Phys. Lett.
B24 (1967) 181.

\bibitem{S} J.J.\ Sakurai, Phys. Lett. 46B (1973) 207.

\bibitem{DAFFR} De Alfaro, S. Fubini, G.\ Furlan and C.\ Rossetti, {\it
Currents
in Hadron Physics} (North-Holland, 1973).

\bibitem{pQCDff}
G.P.\ Lepage and S.J. Brodsky, Phys.\ Lett.\ 87B (1979) 359;
Phys.\ Rev.\ Lett. 43 (1979) 545; A.V.\ Efremov and A.V.\ Radyushkin,
Phys.\ Lett.\ 94B (1980) 245; S.J.\ Brodsky and G.P.\ Lepage,
in {\it Perturabtive Quantum Chromodynamics}, ed.\ A.H.\ Mueller
(World Scientific, Singapore, 1989).

\bibitem{Radyconf} A.V.\ Radyushkin, Acta Phys.\ Polon.\ B15 (1984) 403.

\bibitem{Isgur} N.\ Isgur and C.H.\ Llewellyn Smith,
Phys.\ Rev.\ Lett.\ 52 (1984) 1080; Nucl.\ Phys.\ B317
(1989), 526.

\bibitem{Li} H.-N.\ Li and G.\ Sterman, Nucl.\ Phys.\ B381 (1992) 129.

\bibitem{Akhoury} R.\ Akhoury, Phys.\ Rev.\ D{19} (1979) 1250.

\bibitem{FSZ} G.R.\ Farrar, G.\ Sterman and H.\ Zhang,
Phys.\ Rev.\ Lett.\ {62} (1989) 2229.

\bibitem{Landau} V.B.\ Berestetskii, E.M.\ Lifshitz
and L.P.\ Pitaevskii, {\it Relativistic
Quantum Theory} (Pergamon Press, 1971), p.\ 235.

\bibitem{Eden} R.J.\ Eden, P.V.\ Landshoff, D.I.\ Olive and
J.C.\ Polkinghorne, {\it The Analytic S-matrix} (Cambridge Univ. Press., 1966).

\bibitem{Matveev} V.A.\ Matveev, R.M.\ Muradyan and A.N.\ Tavkhelidze, Lett.
Nuovo Cimento 7 (1973) 719.

\bibitem{Farrar} S.J.\ Brodsky and G.R.\ Farrar, Phys. Rev. Lett. 31
(1973) 1153.

\bibitem{Fey}  R.P.\ Feynman, {\it Photon-Hadron Interactions},
(W.A.\ Benjamin, Reading, MA, 1972), p.\ 145.

\bibitem{CC} C.\ Corian\`{o}, in preparation.

\bibitem{KNizic} E.\ Maina and G.R.\ Farrar, Phys.\ Lett.\ {206B} (1988) 120;
G.
R.\ Farrar and H.\ Zhang, Phys.\ Rev.\ D{41} (1990) 3348;
{42} (1990) 2413(E).

\bibitem{KNizic} A.S.\ Kronfeld and B.\ Ni\v{z}i\'{c}, Phys.\ Rev.\ D{44}
(1991) 3445.

\bibitem{CN} S.\ Coleman and R.E.\ Norton, Nuovo Cimento
Ser.\ 10, 38 (1965) 438.

\bibitem{St77b} G.\ Sterman, Phys.\ Rev.\ D{17} (1978) 2773.

\bibitem{St77a} G.\ Sterman, Phys.\ Lett.\ {73B} (1978) 440.


\end{thebibliography}
\end{document}